\documentclass[10pt,conference]{IEEEtran}
\IEEEoverridecommandlockouts
\def\BibTeX{{\rm B\kern-.05em{\sc i\kern-.025em b}\kern-.08em
    T\kern-.1667em\lower.7ex\hbox{E}\kern-.125emX}}

\newcommand{\tool}{LibFan}
\newcommand{\ourdataset}{LibFanBench}

\newcommand{\stepone}{Class-level Candidate Filtering}
\newcommand{\stepthree}{Method-level Context-Aware Contrastive Learning}
\newcommand{\stepfour}{Functional Partition-based Scoring}

\newcommand{\cgencoder}{Method Call Encoder}
\newcommand{\ctxencoder}{Class Context Encoder}
\newcommand{\crossattn}{Context Fusion Module}

\newcommand{\method}{m}
\newcommand{\methodprime}{m'}
\newcommand{\methoda}{m_a}
\newcommand{\methodb}{m_b}
\newcommand{\methodt}{m_t}

\newcommand{\classa}{c_a}
\newcommand{\classt}{c_t}

\newcommand{\size}[1]{\text{size}(#1)}
\newcommand{\matchset}[1]{\text{Match}_{#1}}

\newcommand{\reach}[2]{\text{R}(#2)}

\newcommand{\cg}{G}
\newcommand{\enhancedcg}{G_{merge}}
\newcommand{\cgdag}{G_{dag}}

\newcommand{\methodsetofnode}[1]{\text{Methods}(#1)}

\newcommand{\ctx}[1]{C_{#1}}
\newcommand{\seq}[1]{T_{#1}}
\newcommand{\embedding}[1]{\mathbf{e}_{#1}}
\newcommand{\class}{c}
\newcommand{\fieldset}[1]{F_{#1}}

\newcommand{\cgvec}{\mathbf{H}_{\text{cr}}}
\newcommand{\ctxvec}{\mathbf{H}_{\text{ctx}}}
\newcommand{\cgmask}{\mathbf{M}_{\text{cr}}}
\newcommand{\ctxmask}{\mathbf{M}_{\text{ctx}}}

\newcommand{\matchmethodset}{M^*}
\newcommand{\methodset}{M}

\newcommand{\partition}{P}
\newcommand{\partitionset}{\mathcal{P}}

\newcommand{\thresholdmethod}{\theta_{m}}
\newcommand{\thresholdpartition}{N_{p}}
\newcommand{\thresholdlib}{\theta_{l}}

\usepackage{listings}
\usepackage{lineno}
\usepackage{float}
\usepackage{booktabs}
\usepackage{pifont}
\usepackage{amssymb}
\usepackage{wrapfig}
\usepackage{amsmath}
\usepackage{multirow}
\usepackage{graphicx}
\usepackage{makecell}
\usepackage{subcaption}
\usepackage{url}
\usepackage{tcolorbox}
\usepackage{xcolor}
\tcbuselibrary{breakable}

\usepackage{multicol}
\usepackage{array}
\usepackage[hidelinks]{hyperref}
\usepackage{minted}

\usepackage[compact]{titlesec}
\usepackage{xcolor}
\usepackage{enumitem}
\usepackage[dvipsnames]{xcolor} %

\usepackage{titlesec}
\titlespacing{\section}{0pt}{0.8ex plus 0ex minus 0.2ex}{0.8ex plus 0ex minus 0.2ex}
\titlespacing{\subsection}{0pt}{0.8ex plus 0ex minus 0.2ex}{0.8ex plus 0ex minus 0.2ex}
\titlespacing{\subsubsection}{0pt}{0.8ex plus 0ex minus 0.2ex}{0.8ex plus 0ex minus 0.2ex}

\setlist[itemize]{leftmargin=1em, itemsep=0pt, topsep=1pt}

\newtcolorbox{rqanswerbox}{
    breakable,           %
    sharp corners,       %
    colback=gray!6,      %
    colframe=black!45,   %
    boxrule=0pt,         %
    leftrule=3pt,        %
    toprule=0pt,         %
    bottomrule=0pt,      %
    rightrule=0pt,       %
    left=6pt,            %
    right=4pt,
    top=4pt,
    bottom=4pt,
    before skip=6pt,
    after skip=6pt
}

\makeatletter
\def\@IEEEaftertitletext{\vspace{-2em}}
\makeatother

\begin{document}

\title{Context-Aware Functional Modeling for Android Third-Party Library Detection}

\author{
\IEEEauthorblockN{
Dihao Fan\IEEEauthorrefmark{1},
Jian Zhang\IEEEauthorrefmark{1},
Yasai Shi\IEEEauthorrefmark{1},
Chuan Luo\IEEEauthorrefmark{1},
Xudong Liu\IEEEauthorrefmark{1},
Yang Liu\IEEEauthorrefmark{2},
Chunming Hu\IEEEauthorrefmark{1},
Xu Wang\IEEEauthorrefmark{1}
}

\IEEEauthorblockA{
\IEEEauthorrefmark{1}Beihang University, Beijing, China\\
\{fandihao, zhangj\_cs, shiyasai, chuanluo, liuxd, hucm, xuwang\}@buaa.edu.cn\\
\IEEEauthorrefmark{2}Nanyang Technological University, Singapore\\
yangliu@ntu.edu.sg
}
}

\maketitle

\begin{abstract}

Third-party libraries (TPLs) are widely used in Android apps, but their reuse can introduce security risks and interfere with downstream program analyses. Existing Android TPL detection approaches face two key limitations: their hand-crafted features are fragile under aggressive code transformations, and their whole-library matching strategies are ineffective when apps retain only part of a TPL. 

In this paper, we propose {\tool}, a learning-based Android TPL detection approach based on context-aware functional modeling. It realizes this modeling through two complementary components: context-aware contrastive learning at the method level and functional partitioning at the library level. At the method level, it learns semantic representations through contrastive training while incorporating outgoing call relationships and class-level context, improving robustness to obfuscation, shrinking, and optimization. At the library level, it partitions each TPL into functionally coherent units and determines library presence using the best-matching partition, thereby accommodating partial library reuse. To evaluate {\tool}, we construct a new benchmark comprising 200 apps and 46 vulnerable TPLs, with each app compiled under four transformation configurations. Under the most challenging R8 full mode, {\tool} achieves F1 scores of 81.3\% at the library level and 47.6\% at the version level, representing relative improvements of 64.9\% and 35.6\% over the state of the art, respectively.

\end{abstract}

\section{Introduction}

The Android operating system hosts one of the largest and most diverse software ecosystems in the world, with over three billion active users across more than 190 countries~\cite{android_wikipedia,business_of_apps2025}. 
A key factor behind Android’s success is its open and modular architecture, which allows developers to rapidly build applications by incorporating a wide range of third-party libraries (TPLs) from the upstream software supply chain. 
Such libraries enable efficient implementation of essential functionalities, including network communication, cloud-based analytics, and crash reporting~\cite{lachgar2018android,shekhar2012adsplit,mahmud2022analysis}.

However, TPL reuse also introduces security and analysis challenges. First, vulnerabilities in TPLs can propagate into downstream apps. For instance, a known \verb|okhttp| vulnerability allows man-in-the-middle attackers to bypass certificate pinning~\cite{okhttp-cve}, and such risks may persist due to complex dependency chains and delayed updates~\cite{derr2017keep,ami2021demo}. Second, accurate TPL identification is also important for Android analyses such as malware classification~\cite{zhang2014semantics,li2024malcertain,yang2024beyond}, app testing~\cite{mahmud2022acid,haryono2021androevolve,wang2024feedback,cao2025intention,hu2023omegatest}, and repackaging detection~\cite{sun2014detecting,zhan2019repackage-study}, where TPL code can otherwise introduce substantial noise.

Therefore, many studies focus on detecting TPLs in Android apps. However, accurate detection is challenged by common code transformations applied during app build and release processes, including \textbf{obfuscation} (e.g., identifier renaming and package flattening), \textbf{shrinking} (e.g., removing unused code), and \textbf{optimization} (e.g., method inlining and class merging). These transformations, introduced by tools such as Android's R8 compiler~\cite{r8}, ProGuard~\cite{ProGuard}, DashO~\cite{DashO}, and Obfuscapk~\cite{Obfuscapk}, can substantially alter the syntactic structure of TPL code, making it difficult to determine whether a given TPL is present in an app~\cite{zhan2020automated,zhan2021research}.

Existing work has proposed various approaches to tackle the above challenges~\cite{li2017libd,orlis,ma2016libradar,zhan2021atvhunter}. For example, LibPecker~\cite{libpecker} extracts class-level features with weighted similarity scoring to tolerate code customization. LibScan~\cite{libscan} adopts a two-stage detection strategy, starting with class-level signature matching followed by method-level static feature comparison; it further enhances method-level matching using call-chain-opcode similarity, achieving strong performance under obfuscation. LibHunter~\cite{libhunter} takes a step further by optimizing both class-level and method-level feature design to specifically counter code optimization techniques such as call site optimization and method inlining, thereby improving TPL detection performance in optimization-heavy scenarios.

Despite recent advances, two critical challenges in Android TPL detection remain underexplored:
\begin{itemize}
\item \textbf{Fragility of Hand-crafted Features.}
Existing approaches largely rely on hand-crafted instruction-level features, using opcodes as the primary signal for method matching, sometimes supplemented by cues such as string literals.
Although effective against common obfuscation, such features remain fragile under aggressive transformations, particularly compiler optimizations that rewrite instructions while preserving functionality.
This fragility can impair method matching and propagate errors to library-level detection.
While prior work attempts to tolerate such variations through heuristics such as incorporating call-chain opcodes, they remain fundamentally tied to low-level instruction patterns and do not learn high-level functional representations that remain stable across transformations.
\item \textbf{Partial TPL Presence in Apps.}
In practice, apps often reuse only a subset of a TPL's functionality.
Our empirical analysis reveals that, after aggressive shrinking and optimization in release builds, only 34.6\% of a TPL's classes and 16.7\% of its methods remain in the final app on average.
Still, most existing tools treat each TPL as a complete unit and aggregate matching evidence across all of its classes or methods.
Consequently, even when the retained components are correctly matched, their evidence can be overwhelmed by the large number of unretained components, underestimating the overall app-TPL similarity and ultimately causing missed detections.

\end{itemize}

In this paper, we propose \textbf{\underline{Lib}}rary \textbf{\underline{F}}inder for \textbf{\underline{An}}droid, namely {\tool}, a learning-based TPL detection tool based on context-aware functional modeling. {\tool} follows a three-stage workflow: class-level candidate filtering, context-aware method matching, and functional partition-based library scoring. It adopts the candidate-filtering strategy from prior work and introduces the latter two stages to model TPL functionality at the method and library levels, respectively.

Given the resulting candidate class pairs, {\tool} performs method-level matching using a dual-encoder framework, instead of relying on fragile hand-crafted method features. Each method is independently encoded into a semantic representation using its decompiled Java source code together with two complementary forms of context: (1) a call context summarizing its outgoing calls and (2) a class context providing selectively chosen structural cues relevant to its functionality. The encoders are initialized from a pretrained code language model and fine-tuned using a contrastive objective constructed from transformed and non-transformed method pairs. This objective brings functionally equivalent methods closer in the embedding space while separating non-equivalent ones, enabling more robust matching under instruction-level rewriting introduced by obfuscation and compiler optimization.

At the library level, {\tool} introduces a functional partition-based scoring strategy to address partial TPL presence caused by shrinking and optimization. Unlike existing approaches that treat a TPL as a single matching unit, it groups the library's methods into functionally coherent partitions according to their internal call relations. Each partition is initialized from a public entry point and its reachable methods, and overly small partitions are iteratively merged to avoid excessive fragmentation. {\tool} then computes a match ratio for each partition independently and uses the highest ratio as the final app-TPL similarity. In this way, a strongly matched retained functionality can determine library presence without being diluted by other library components removed during compilation. 

To evaluate {\tool}, we construct a new benchmark, {\ourdataset}, since we identified that existing datasets often use only explicitly declared build-script dependencies as ground truth and overlook transitive dependencies introduced during compilation, leading to incomplete or inaccurate labels.
{\ourdataset} comprises 200 open-source apps and 46 vulnerable TPLs. To assess robustness under increasingly aggressive transformations, each app is compiled into four variants using the D8/R8 toolchain: D8 without transformation, R8 with shrinking, R8 with optimization and shrinking, and R8 full mode with optimization, obfuscation, and shrinking. Ground-truth TPLs are identified from compilation artifacts, including the \textit{sdk-dependencies} directory, yielding more complete and realistic labels.
On this benchmark, {\tool} achieves an F1 score of 81.3\% at the library level and 47.6\% at the version level on the most challenging variant (R8 full mode), representing relative improvements of 64.9\% and 35.6\% over the state-of-the-art, respectively.

Our contributions are summarized as follows:
\begin{itemize}
\item We propose a context-aware contrastive learning framework for Android TPL detection. By jointly encoding a method's call relationship and class context, the fine-tuned dual-encoder model learns semantic method representations that remain robust under code transformations.
\item We design a functional partition-based similarity strategy to address missed detections caused by partial TPL reuse. Instead of aggregating matches over the entire library, {\tool} scores the most relevant functional partition, improving robustness under code shrinking and optimization.
\item We construct a more realistic benchmark comprising 200 apps and 46 vulnerable TPLs, accounting for transitive dependencies and four transformation configurations. On this benchmark, {\tool} achieves 81.3\% library-level and 47.6\% version-level F1 scores under the most challenging R8 full mode, substantially outperforming existing approaches.
\end{itemize}

\section{Motivation}\label{background}

\textbf{Background.} In the Android build pipeline, D8 translates Java \textit{.class} bytecode into Dalvik \textit{.dex} format, while R8 additionally performs shrinking, optimization, and obfuscation for release builds. As a result, Android TPL detection commonly needs to compare transformed app code against untransformed reference TPLs. Existing tools typically follow a staged matching pipeline: they first filter candidate class pairs, then match methods using identifier-agnostic hand-crafted features such as opcode sequences or string literals, and finally aggregate matched methods over the whole TPL. However, this design still has two fundamental limitations under R8 transformations.

\begin{figure}[htbp]
\vspace{-0.5em}
  \centering
  \includegraphics[width=1\linewidth]{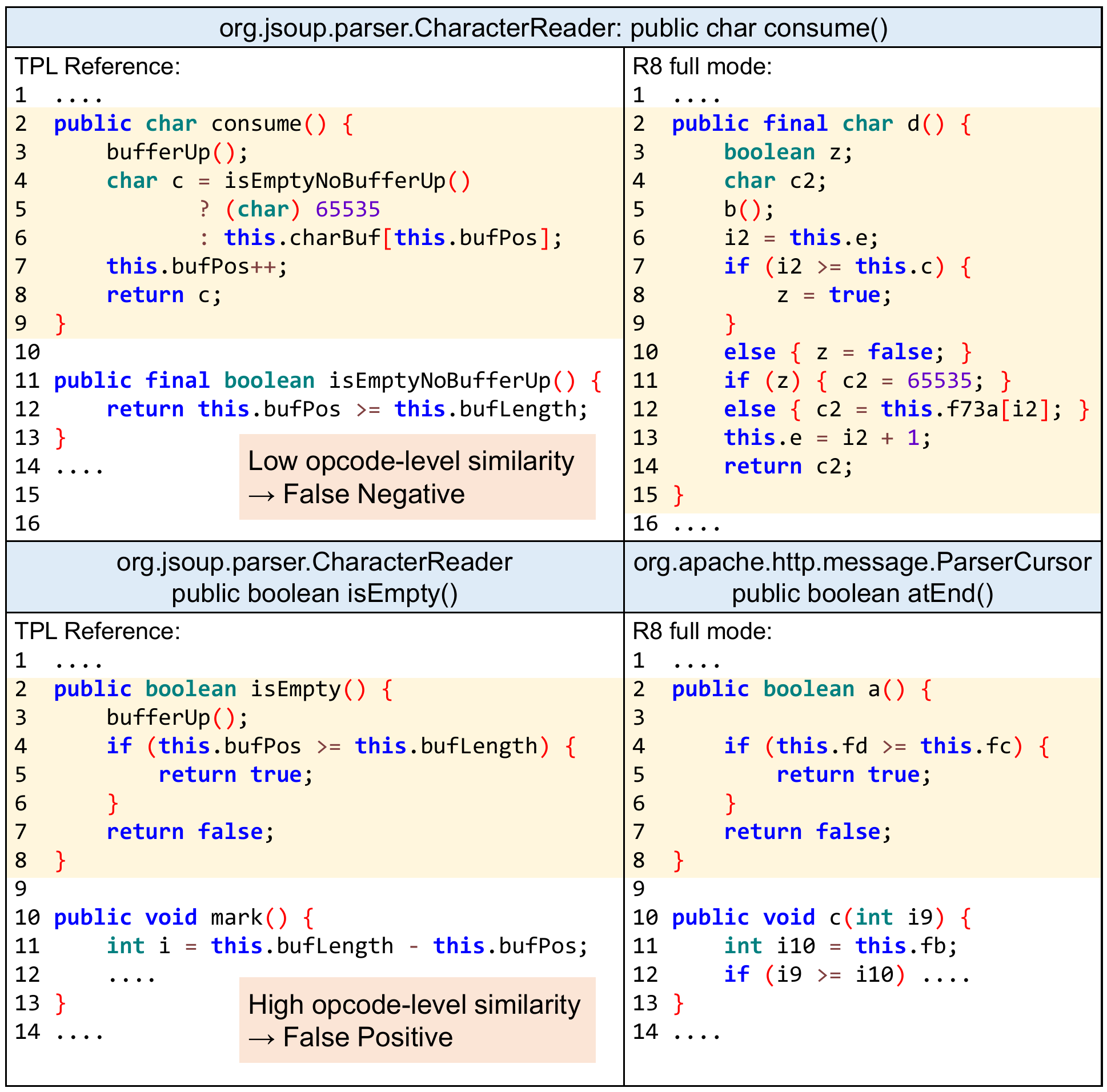}
  \caption{Motivating examples of opcode-level method matching failures under R8 transformations.}
  \label{fig:motivation}
  \vspace{-0.5em}
\end{figure}

\textbf{L1: Method-level hand-crafted feature matching is unreliable and ambiguous under R8.}
Existing tools commonly rely on opcode-level features for method matching~\cite{libscan,libhunter}, but such features do not reliably capture high-level functional semantics. Figure~\ref{fig:motivation} illustrates two representative failure cases, where similarity is quantified as the Jaccard similarity~\cite{jaccard-similarity} over the set of opcode types extracted from each method.

In the first case, the callee \texttt{isEmptyNoBufferUp} is inlined into \texttt{consume}, substantially changing the opcode composition of the resulting transformed method \texttt{d}. As a result, the similarity between the reference method and its R8-transformed counterpart drops to only 52.6\%, even though they are semantically equivalent, leading to a false negative.

In the second case, \texttt{isEmpty} and \texttt{atEnd} operate over different class states and participate in different surrounding behaviors, making them semantically mismatched. However, opcode-level matching reports a high similarity of 87.5\% between the two methods, leading to a false positive. 

These examples show that isolated opcode-based method features are insufficient for robust method-level matching. Accurate matching requires contextual semantics from both call relationships and the enclosing class, motivating our context-aware method representation. In the two cases shown in Figure~\ref{fig:motivation}, {\tool} assigns a high score of 90.2\% to the true match and a low score of 35.2\% to the semantic mismatch.

\textbf{L2: Library-level similarity estimation over the full method set is ill-posed under partial TPL reuse.} Existing approaches typically estimate app--TPL similarity by normalizing matched methods against the entire method set of the reference TPL. This design implicitly assumes that most TPL code is retained in the app, but the assumption breaks under R8 shrinking and optimization, which remove unused library code. In the training and validation apps of our benchmark, only 48.9\% of classes and 37.7\% of methods per TPL are retained in the final app on average after code shrinking; with further optimization, these ratios drop to 34.6\% and 16.7\%, respectively. These findings indicate that apps often reuse only a fraction of a TPL’s functionality. Consequently, even when all retained methods are correctly matched, whole-library scoring can still produce a low similarity score because the removed methods dominate the denominator, leading to false negatives.

These limitations motivate context-aware functional modeling along two complementary directions: (1) method matching should capture contextual semantics beyond isolated opcode-based features, and (2) library scoring should focus on the functional portions of a TPL retained in the app rather than on the entire library.

\section{Approach}

In this section, we present our learning-based TPL detection tool {\tool}, which introduces context-aware contrastive learning to robustly encode method functionality, and functional partitioning to focus matching on the most relevant subset of each TPL.

\begin{figure*}[htbp]
  \centering
  \includegraphics[width=0.82\textwidth]{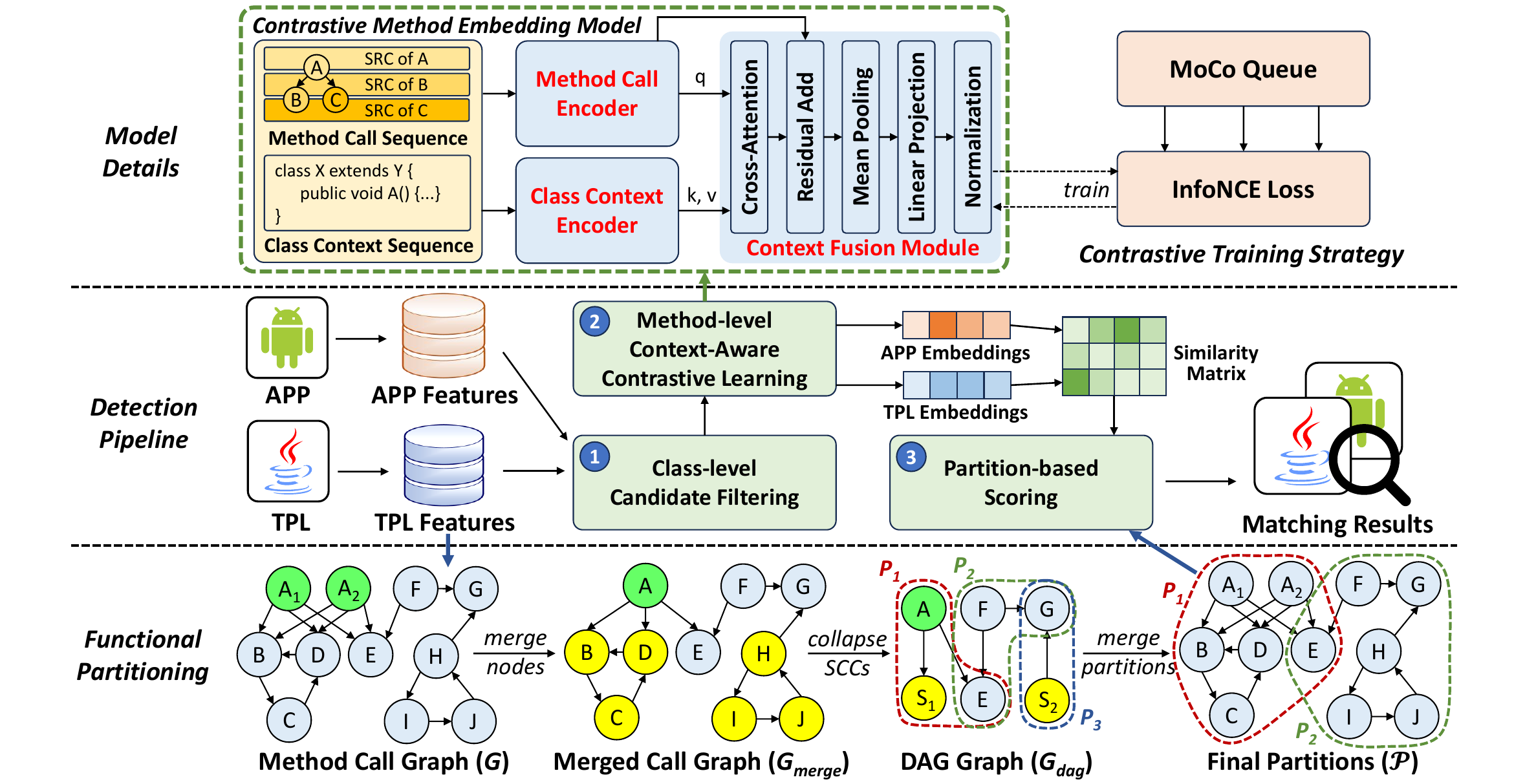}
  \caption{The overall workflow of \tool}
  \label{fig:workflow}
  \vspace{-2em}
\end{figure*}

Figure~\ref{fig:workflow} illustrates the overall workflow of {\tool}.
The middle part shows the detection pipeline, which takes an app and a candidate TPL as input and outputs whether the TPL is reused in the app, consisting of three main stages:

\noindent \ding{182} \textbf{\stepone} (Section~\ref{sec:step1}): 
{\tool} performs class-level pre-matching to filter out irrelevant app--TPL class pairs, thereby reducing the search space for subsequent fine-grained matching.

\noindent \ding{183} \textbf{\stepthree} (Section~\ref{sec:step3}):
For each method in the retained app--TPL class pairs, {\tool} applies a context-aware contrastive embedding model to derive method representations and compute method-level semantic similarity.

\noindent \ding{184} \textbf{\stepfour} (Section~\ref{sec:step4}):
{\tool} aggregates method-level matching evidence into a library-level reuse score via functional partitioning, and determines TPL reuse based on this score.

\subsection{\stepone}\label{sec:step1}

Initially, {\tool} reuses LibHunter's signature-based class matching component~\cite{libhunter} as a coarse-grained filter to prune irrelevant app--TPL class pairs.
Each class is represented by an obfuscation-resilient signature consisting of field type descriptors and method prototypes. 
An app--TPL class pair is retained when the app class signature is covered by the TPL class signature, and the retained pairs are forwarded to subsequent method-level matching.

\subsection{\stepthree}\label{sec:step3}

As discussed in Section~\ref{background}, existing Android TPL detection tools primarily rely on hand-crafted method-level features and fail to capture high-level functional semantics, resulting in limited robustness under code obfuscation and optimization.
To address this limitation, we propose a representation learning-based approach for Android TPL detection.
We formulate the matching task at the method level, as methods constitute the fundamental units of functionality in both apps and TPLs.
Unlike prior approaches, we decompile binaries into Java source code and embed methods in a context-aware manner by jointly considering their call relationships and enclosing class context.
We train the model using a contrastive learning strategy (Section~\ref{subsec:train}) to ensure that methods with equivalent high-level semantics are mapped close to each other in the embedding space, even under substantial compiler-induced transformations.

The upper left part of Figure~\ref{fig:workflow} illustrates the architecture of our proposed Contrastive Method Embedding Model.
Given a target method $\method$, we encode it from two perspectives. The \textbf{\cgencoder} (Section~\ref{subsec:cg-encoder}) captures its call context by jointly encoding $\method$ and its downstream reachable methods, while the \textbf{\ctxencoder} (Section~\ref{subsec:ctx-encoder}) captures its class context through a focused representation of the enclosing class.
 By default, we adopt the encoder-only version of UniXcoder~\cite{unixcoder} as the base model for both encoders, since it is lightweight and demonstrates strong performance in tasks based on code understanding~\cite{wang2025element-repair,shi2023cocosoda,liu2025uncertainty-search}.
The outputs of the two encoders are fused via a \textbf{\crossattn} (Section~\ref{subsec:cross-attn}).
We employ a cross-attention mechanism to integrate the auxiliary class context into the method representation. The fused features are aggregated to yield the final fixed-length method embedding $\embedding{\method}$.

\subsubsection{\cgencoder}
\label{subsec:cg-encoder}

As noted in~\cite{libhunter}, the Android R8 compiler performs method inlining during the build process, by merging callers and callees and interleaving semantics across methods.
Consequently, method-level representations that ignore call relationships are inherently fragile under such optimizations.
This motivates the incorporation of inter-procedural context to capture method semantics more robustly.

We enhance the representation of a target method $\method$ by incorporating all downstream methods reachable from $\method$ in the program call graph $\cg$, which together capture richer method-level semantics. We denote this set of reachable methods, together with $\method$ itself, as $\reach{\cg}{\method}$.
Unlike LibHunter~\cite{libhunter}, which selectively constructs call chains based on R8’s inlining strategy, we do not attempt to statically infer inlining behavior. In practice, code shrinking alters the number of call sites and thus affects inlining decisions, making inference based solely on the reference TPL unreliable.

Since pretrained language models operate on sequential inputs, we linearize the decompiled Java source code of the methods in $\reach{\cg}{\method}$. We adopt a breadth-first traversal (BFS) over the call subgraph rooted at $\method$ as a practical default. BFS preserves semantic locality by placing methods closer to $\method$ earlier in the sequence. This ordering aligns well with inlining effects, as $\method$ and its direct callees tend to remain in close proximity when flattened into an optimized method $\methodprime$. The token sequences of the visited methods are concatenated in traversal order to form the input sequence $\seq{\method}$ for {\cgencoder}. The encoder then embeds $\seq{\method}$ into a feature representation $\cgvec$, while retaining the valid token mask $\cgmask$ of $\seq{\method}$ for subsequent aggregation.

\subsubsection{\ctxencoder}\label{subsec:ctx-encoder}

\begin{figure}[t]
\vspace{0pt}
    \raggedright
    \includegraphics[width=0.85\linewidth]{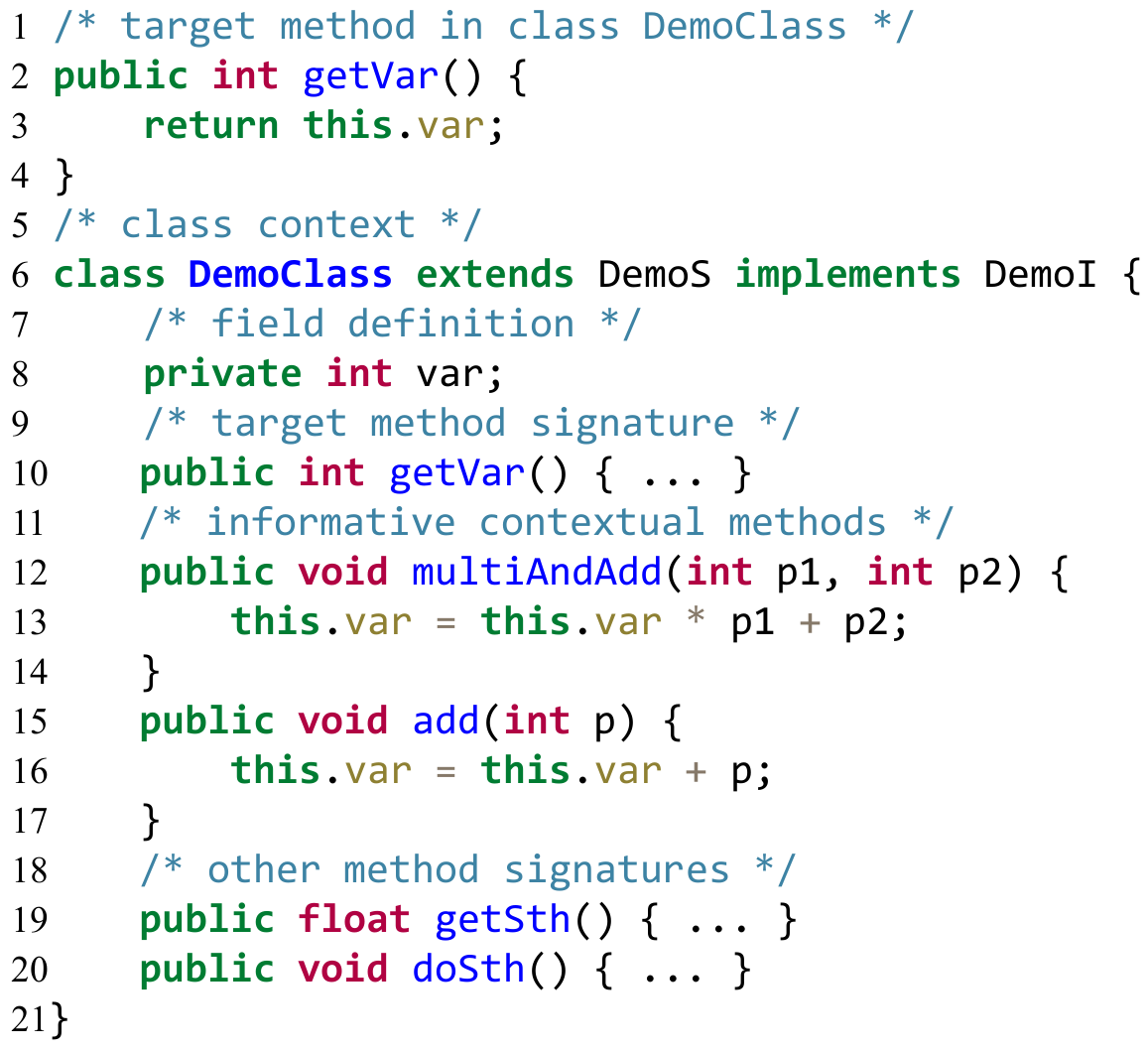}
    \caption{An example of the class context used for the \ctxencoder.}
    \vspace{-0.9em}
    \label{fig:ctx}
\end{figure}

Although the {\cgencoder} captures rich method-level semantics through call relationships, it may provide limited discrimination for methods with weak or no call dependencies, such as structurally isolated or semantically generic methods. To complement such cases, we introduce the {\ctxencoder}, which encodes a focused class-level context for each target method $\method$. As illustrated in Figure~\ref{fig:ctx}, for the target method \verb|getVar|, the context sequence starts with the class header, including the class name, inheritance hierarchy, and implemented interfaces.

Due to the limited input length of the encoder and the code removal introduced by shrinking, LibFan avoids encoding the entire enclosing class $\class$. Instead, it first identifies the set of fields $\fieldset{\method}$ accessed by $\method$ and constructs the class context $\ctx{\method}$ around these fields. In Figure~\ref{fig:ctx}, \verb|getVar| accesses the field \verb|var|, i.e., $\fieldset{\texttt{getVar}}=\{\texttt{var}\}$. The resulting context includes the declarations of fields in $\fieldset{\method}$, the bodies of methods that access these fields, and the signatures of the target method and the remaining methods in the class. For example, the declaration of \verb|var| and the bodies of \verb|multiAndAdd| and \verb|add| are retained because they share the accessed field with \verb|getVar|, while unrelated methods are kept only as signatures to preserve class structure. The target method itself is represented only by its signature, since its body has already been encoded by the {\cgencoder}.

This design captures field-relevant class semantics while avoiding redundant or unstable class-wide code. Finally, the {\ctxencoder} encodes $\ctx{\method}$ and outputs the feature sequence $\ctxvec$ together with the valid-token mask $\ctxmask$.

\subsubsection{\crossattn}\label{subsec:cross-attn}
We design the {\crossattn} to fuse the features, treating the class context as auxiliary information to enhance the method-level semantic representation.
Specifically, we utilize Multi-Head Attention (MHA) to perform a cross-attention mechanism. The output sequence $\cgvec$ of the {\cgencoder} serves as the query because method matching is the primary task, and $\ctxvec$ of the {\ctxencoder} serves as both key and value, acting as an auxiliary context memory from which the method representation retrieves relevant class-level cues. The multi-head mechanism allows the model to capture diverse types of contextual dependencies from this memory simultaneously. The mask $\ctxmask$ is applied to exclude invalid tokens in the context. 
To integrate class-level context without overwhelming method semantics, we add a residual connection from $\cgvec$ to the attention output, forming the fused representation $\mathbf{H}_{\text{fused}}$.
\begin{equation}
    \mathbf{H}_{\text{fused}}=\cgvec + \text{MHA}(\cgvec, \ctxvec, \ctxvec, \ctxmask)
\end{equation}
$\mathbf{H}_{\text{fused}}$ is then aggregated via masked mean pooling using $\cgmask$ to consider only valid tokens.
Finally, a linear projection $\mathbf{W}_p$ followed by L2 normalization yields the final embedding $\embedding{\method}$.
\begin{equation}
    \mathbf{e}_m = \text{L2Norm}\left( \mathbf{W}_p \cdot \text{MeanPool}(\mathbf{H}_{\text{fused}}, \cgmask) \right)
    \label{eq:embedding}
\end{equation}

For two methods $\methoda$ and $\methodb$, we compute their similarity score using cosine similarity~\cite{cosine-similarity}. Since the two embeddings have been normalized, the similarity score reduces to a simple dot product:
\begin{equation}
   \text{cos\_sim}(\embedding{\methoda},\embedding{\methodb})=\embedding{\methoda}\cdot\embedding{\methodb}
\label{eq:cos}
\end{equation}

\subsubsection{Model Training}\label{subsec:train}

Pretrained code models such as UniXcoder are not explicitly trained to recognize the same method after obfuscation and compiler optimization substantially rewrite its code. Consequently, transformation-equivalent methods may receive dissimilar representations. To address this challenge, we fine-tune our method embedding model through contrastive learning. The key to our training design is the construction of transformation-aware method pairs, which directly teach the model to preserve functional similarity across code transformations.

Specifically, we compile each training app into two variants: an R8 variant with shrinking, optimization, and obfuscation enabled, and a corresponding D8 variant without these transformations. Each method retained in the R8 variant is paired with its original D8 counterpart to form a positive training pair. We establish method correspondence using the mapping file generated during compilation. Because shrinking and method inlining may remove or merge methods in the R8 output, we retain only methods that remain individually identifiable in both variants. To increase training diversity, we use all eligible app methods rather than restricting the data to TPL methods. We further deduplicate the resulting pairs according to their method signatures in the D8 variant, including the class name, method name, return type, and parameter list, retaining one pair for each unique signature.

During training, each R8-compiled method serves as a query, and its corresponding D8-compiled method serves as the positive sample. The D8 methods associated with other queries in the same mini-batch are used as in-batch negatives. We further adopt Momentum Contrast (MoCo)~\cite{he2020moco}, which maintains a queue of D8 method embeddings generated in previous iterations. Thus, the negative set for each query consists of non-corresponding D8 methods from both the current mini-batch and the MoCo queue.

We employ the InfoNCE loss~\cite{oord2018infonce} for optimization:
\begin{equation}
\label{eq:infonce}
\mathcal{L}_{\text{InfoNCE}} = - \log \frac{\exp({\text{cos\_sim}(\mathbf{e}, \mathbf{e^+}) / \tau})}{\sum\limits_{\mathbf{e'} \in \mathcal{E}} \exp({\text{cos\_sim}(\mathbf{e}, \mathbf{e'}) / \tau})},
\end{equation}
where $\mathbf{e}$ and $\mathbf{e^+}$ denote the embeddings of the R8-compiled query method and its corresponding D8-compiled positive method, respectively, as obtained from Equation~\ref{eq:embedding}. $\mathcal{E}$ contains the positive sample, the non-corresponding D8 methods in the current mini-batch, and the queued samples from previous iterations, while $\tau$ is a temperature hyperparameter. This objective increases the similarity between each transformed method and its original counterpart while reducing its similarity to non-corresponding methods. Through this transformation-aware supervision, the model learns method representations that are less sensitive to code rewriting introduced by obfuscation and compiler optimization.

\subsection{\stepfour}\label{sec:step4}
As discussed in Section~\ref{background}, whole-library similarity estimation is ill-suited to partial TPL reuse, because code shrinking and optimization may remove large portions of unused library code.
To address this limitation, {\tool} scores TPL reuse at the functional-partition level rather than over the entire library.
It partitions each TPL into functionally coherent method sets and uses the best-matching partition to determine library-level presence.

\subsubsection{Functional Partitioning for TPL}\label{subsec:tpl-partition}

We start from the observation that TPL functionality is commonly exposed through public entry methods~\cite{wang2020tplstudy}, whose reachable call subgraphs approximate reusable functional units. However, directly using these subgraphs may produce overly small partitions, making the score sensitive to individual false-positive method matches.

To avoid overly fine-grained partitions, we design a functional partition merging strategy that merges partitions smaller than $\thresholdpartition$ whenever possible.
As a preprocessing step, we merge overloaded methods in the same class, since they usually serve as alternative entry points to the same functionality. Specifically, for methods that share the same name within the same class in the TPL global call graph $\cg$, we create a new overload-group node and redirect all incoming and outgoing edges of the original nodes to this new node. This produces a merged call graph $\enhancedcg$ that retains call dependencies while reducing unnecessary graph fragmentation.

Furthermore, many public methods in a TPL are invoked internally by other methods within the same library. This implies that the rooted subgraph of a node $v_a$ in $\enhancedcg$, which has predecessors in the call graph, is always a subgraph of the rooted subgraph of $v_b$ with in-degree zero. The union of the subgraphs rooted at in-degree-zero methods can cover all functional logic within the TPL.
However, this assumption breaks down in the presence of cyclic method calls in Java where no single method in the cycle has in-degree zero. These cycles form strongly connected components (SCCs) in the call graph. To handle such cases, we first extract SCCs and collapse each SCC into a single node, thereby converting the merged call graph $\enhancedcg$ into a directed acyclic graph (DAG), denoted as $\cgdag$, where each node $v$ represents a set of methods $\methodsetofnode{v}$ instead of a single one.
This transformation guarantees that every node in $\cgdag$ is included in the subgraphs rooted at the in-degree-zero SCCs, which serve as the base units for subsequent functional partitioning. 

Based on $\cgdag$, we construct the initial functional partition set $\partitionset$ as the unions of methods reachable from each node with zero in-degree in the node set $V_{\cgdag}$. %
\begin{gather}
\partitionset = \left\{ \bigcup\nolimits_{u \in \reach{\cgdag}{v}}u \;\middle|\; v \in V_{\cgdag},\ \text{in-degree}(v) = 0 \right\}
\label{eq:partition}
\end{gather}

We then iteratively refine $\partitionset$ to ensure that each partition contains a sufficient number of methods. In each iteration, we select the smallest partition $\partition^* \in \partitionset$ whose size is below the threshold $\thresholdpartition$. For $\partition^*$, we compute the Jaccard similarity~\cite{jaccard-similarity} with every other partition $\partition \in \partitionset$, and merge it with the partition that yields the highest similarity.
If multiple candidate partitions achieve the same similarity score, 
we break ties by comparing the Jaccard similarity of their associated class sets and package sets, and merge $\partition^*$ with the most similar one. If no preference can be established after these comparisons, we merge $\partition^*$ with the second smallest partition in $\partitionset$. This process is repeated until $\partition^*$ exceeds the size threshold $\thresholdpartition$ or $\partitionset$ contains only a single remaining partition. 
\begin{equation}
    \text{Jaccard\_sim}(P_1,\ P_2)=\frac{|P_1\ \cap\ P_2|}{|P_1\ \cup\ P_2|}
    \vspace{-0.2em}
\end{equation}

The lower part of Figure~\ref{fig:workflow} summarizes this process: overloaded methods are first merged, SCCs are collapsed into a DAG, and small reachable method sets are further merged into the final functional partitions.

\subsubsection{TPL and version scoring}\label{subsec:tpl-score}

In the final scoring stage, we first perform filtering based on the cosine similarity between embedding $\embedding{\methoda}$ from the app method $\methoda$ and embedding $\embedding{\methodt}$ from the TPL method $\methodt$. We discard method pairs with similarity lower than $\thresholdmethod$.

Let $\matchset{\classa,\classt}$ denote the set of matched method pairs between a candidate app class $\classa$ and a candidate TPL class $\classt$ that survive the class-level filtering stage. Let $\size{\methodt}$ denote the number of opcode instructions in the compiled body of method $\methodt$. %
We compute the confidence score between each pair of classes as:
\begin{equation}
    \text{Confidence}(\classa,\classt)=\sum\nolimits_{(\methoda,\methodt)\in \matchset{\classa,\classt}}\size{\methodt}
\end{equation}

For each TPL class $\classt$, we select its matched app class $\classa^*$ with the highest confidence score, and preserve only methods of $\classt$ in $\matchset{\classa^*,\classt}$. For each TPL, we then take the union of the preserved matched methods $\matchmethodset$ across all its classes.

For \textbf{library-level detection}, we first compute the ratio of matched methods within each functional partition $\partition \in \partitionset$. %
We then define the library-level similarity score between the app and the target TPL as the maximum partition match ratio:
\begin{gather}
    \text{Lsim} = \max_{\partition \in \partitionset}{\frac{\sum_{m\in {\partition\cap \matchmethodset}}{\size{m}}}{\sum_{m\in \partition}{\size{m}}}}
\end{gather}

For \textbf{version-level detection}, we revert to the global similarity computation where $\methodset$ denotes all methods in the TPL, as differences across versions may not necessarily correlate with the most similar partition:
\begin{equation}\label{eq:vsim}
    \text{Vsim}=\frac{\sum\nolimits_{m\in M^*}{\size{m}}}{\sum\nolimits_{m\in \methodset}{\size{m}}}
\end{equation}

We consider the app to contain the TPL when $\text{Lsim}$ exceeds the threshold $\thresholdlib$. Once the presence of the TPL is confirmed, we select the version with the highest $\text{Vsim}$ among all versions of the TPL as our version result.

\section{Evaluation}\label{evaluation}

In this section, we conduct experiments to evaluate our proposed tool {\tool} on Android TPL detection tasks by answering the following research questions:

\begin{itemize}
    \item \textbf{RQ1:} How does {\tool} perform compared with state-of-the-art Android TPL detection tools, and is it practical in runtime? (Effectiveness and Efficiency)
    \item \textbf{RQ2:} How do different components of {\tool} and the choice of pretrained representation model affect the final performance? (Ablation Study)
    \item \textbf{RQ3:} Is the performance of {\tool}'s functional partitioning sensitive to the minimum number of methods $\thresholdpartition$ required in each partition? (Sensitivity)
    \item \textbf{RQ4:} How does {\tool} perform on unseen TPLs and in-the-wild apps? (Generalizability)
\end{itemize}

\subsection{Experiment Setup}\label{sec:dataset}

\subsubsection{Baselines} We select three state-of-the-art tools as baselines for evaluating effectiveness, including LibHunter~\cite{libhunter}, LibScan~\cite{libscan}, and LibPecker~\cite{libpecker}. We do not include other tools, as they were not designed to handle complex R8 code transformation scenarios, and prior work~\cite{libhunter,libscan} has demonstrated their limited effectiveness under such conditions.

\subsubsection{Dataset}
Although existing work~\cite{libhunter,libscan} has prepared large-scale datasets targeting R8-compiled apps with code shrinking and code optimization, we find that the ground truth in these datasets is often incomplete. In particular, they extract the TPLs used by an app based on the Gradle configuration of the source project, without considering transitive dependencies introduced during compilation. This leads to two main issues:
(1) Tools may correctly detect transitive dependencies (e.g., \texttt{okhttp} via \texttt{retrofit}), but since these are not explicitly declared in Gradle, evaluations mislabel them as false positives.
(2) Transitive dependencies that contain critical known vulnerabilities (e.g., \verb|okio|~\cite{okio-cve}) are omitted from previous datasets. %
Therefore, we construct a dataset, namely \ourdataset, with more complete and realistic ground truth that better reflects real-world library usage patterns.

We collected 200 Android app projects from F-Droid~\cite{f-droid} whose Gradle builds generate the \textit{sdk-dependencies} folder, which provides complete dependency information under AGP 4.0.0 or later. To evaluate robustness under different R8 transformations, we compiled each app into four variants with progressively stronger transformations: \textbf{D8} (non-transformed), \textbf{Srk} (shrinking enabled), \textbf{Opt+Srk} (optimization and shrinking enabled), and \textbf{Opt+Obf+Srk} (R8 full mode with optimization, obfuscation, and shrinking all enabled). By analyzing the \textit{sdk-dependencies} and referencing Maven~\cite{maven}, we identified 46 TPL artifacts, each with at least one known vulnerability, covering a total of 3,120 versions. To ensure fair comparison, following LibScan, we primarily retained versions actually used in the app set. In addition, we addressed TPLs with sparse usage by including adjacent versions and ensuring at least one vulnerable instance per TPL artifact. The final set comprises 369 versions, with each TPL represented by at least five distinct versions. 
We further used the post-compilation \textit{usage.txt} file to remove dependencies eliminated by R8 shrinking or optimization from the ground truth. The resulting dataset contains 831, 757, 749, and 749 app--TPL pairs for D8, Srk, Opt+Srk, and Opt+Obf+Srk, respectively.

\subsubsection{Metrics}
Following existing work, we evaluate all tools at both the library and version levels. At the \textbf{library level}, a detection is true positive (TP) if the reported TPL is actually used by the app; otherwise, missing used TPLs and reporting unused TPLs are counted as false negative (FN) and false positive (FP), respectively. At the \textbf{version level}, library-level FPs and FNs are inherited. For each library-level TP, the result is counted as a version-level TP only if the reported version matches the ground truth; otherwise, it is counted as a version-level FP. We report precision, recall, and F1 score as the results.

\subsection{Implementation}\label{sec:implementation}

All experiments were conducted on a machine equipped with Intel(R) Xeon(R) Platinum 8358 CPUs and an NVIDIA A100 GPU, running Ubuntu 24.04.1 LTS. We converted all TPL \textit{.jar} and \textit{.aar} files into \textit{.dex} format using D8, and used JADX 1.4.7~\cite{Jadx} to decompile bytecode into Java source code. Following LibHunter, we used Androguard 3.3.5~\cite{Androguard} to extract features, construct call graphs and build TPL partitions. We reused the signature-based class matching module from the released LibHunter implementation. For methods that JADX failed to decompile, we fell back to Androguard to obtain their source-like representations.
The maximum input length of UniXcoder was set to 1,024 tokens, with other hyperparameters following UniXcoder-base~\cite{unixcoder-base}.

We randomly split {\ourdataset} by apps into training, validation, and testing sets with a ratio of 10\%, 10\%, and 80\%, respectively. The training set contains 55,406 method pairs for fine-tuning, while the validation set uses 6,880 R8-compiled TPL methods as queries and 37,628 D8-compiled app methods as candidates. The checkpoint with the highest MRR~\cite{mrr} is selected for hyperparameter tuning and final evaluation.

We tune $\thresholdmethod$, $\thresholdlib$, and $\thresholdpartition$ by grid search on the validation set, with $\thresholdmethod,\thresholdlib \in [0,1]$ at a step of 0.05 and $\thresholdpartition \in [100,1000]$ at a step of 100. The selected settings are $\thresholdpartition=600$, $(\thresholdmethod,\thresholdlib)=(0.80,0.40)$ for D8 and $(0.90,0.05)$ for R8 variants, following LibScan~\cite{libscan}. Baseline hyperparameters are tuned on the same validation set for fairness.

\subsection{Effectiveness and Efficiency (RQ1)}

\begin{table}[htbp]
    \renewcommand{\arraystretch}{1.08}
    \setlength{\extrarowheight}{1pt}
    \caption{The overall effectiveness and efficiency of {\tool} and baseline tools. Precision (P), Recall (R), and F1 are shown as library-level / version-level percentages, while T reports the end-to-end online inference time amortized per app--TPL pair. Bold indicates the best result.}
    \label{tab:tpl-comparison}
    \vspace{-0.5em}
    \centering
    \resizebox{\columnwidth}{!}{
    \begin{tabular}{c|c|c|c|c|@{\hspace{2pt}}c@{\hspace{2pt}}}
        \specialrule{1pt}{0pt}{0pt}
        Tool & M. & D8 & Srk & Opt+Srk & Opt+Obf+Srk \\
        \hline
        \multirow{4}{*}{LibPecker}
& P  & 93.9 / 92.9 & 95.5 / 67.0 & 100 / 100 & 100 / 100 \\
& R  & 71.6 / 71.4 & 28.3 / 21.7 & 0.8 / 0.8 & 0.8 / 0.8 \\
& F1 & 81.3 / 80.8 & 43.7 / 32.8 & 1.7 / 1.7 & 1.7 / 1.7 \\
\cline{2-6}
& T  & 23.084s & 11.770s & 8.433s & 8.273s \\

        \hline
        \hline
        \multirow{4}{*}{LibScan}
& P  & 93.6 / 88.8 & 88.6 / 65.1 & 91.5 / 39.0 & 91.4 / 36.2 \\
& R  & 82.9 / 82.1 & 42.4 / 35.1 & 9.0 / 4.1 & 8.9 / 3.7 \\
& F1 & 87.9 / 85.3 & 57.3 / 45.6 & 16.5 / 7.4 & 16.2 / 6.7 \\
        \cline{2-6}
& T  & \textbf{0.560s} & \textbf{0.203s} & \textbf{0.184s} & \textbf{0.172s} \\

        \hline
        \hline
        \multirow{4}{*}{LibHunter}
& P  & 93.4 / 86.2 & 58.3 / 46.7 & 68.2 / 44.1 & 69.2 / 45.0 \\
& R  & 89.2 / 88.4 & 66.6 / 61.5 & 38.0 / 28.4 & 38.3 / 28.8 \\
& F1 & 91.2 / 87.3 & 62.1 / 53.0 & 48.8 / 34.6 & 49.3 / 35.1 \\
\cline{2-6}
& T  & 2.507s & 0.782s & 0.474s & 0.470s \\

        \hline
        \hline
        \multirow{4}{*}{\tool}
& P  & 89.0 / 77.7 & 81.3 / 52.9 & 85.6 / 41.0 & 86.1 / 39.3 \\
& R  & 100 / 100 & 91.4 / 87.3 & 79.4 / 64.9 & 76.9 / 60.3 \\
& F1 & \textbf{94.2} / \textbf{87.4} & \textbf{86.0} / \textbf{65.9} & \textbf{82.4} / \textbf{50.2} & \textbf{81.3} / \textbf{47.6} \\
\cline{2-6}
& T  & 2.193s & 0.897s & 0.642s & 0.644s \\

        \specialrule{1pt}{0pt}{0pt}
    \end{tabular}}
    
\par\vspace{0.3em}
\raggedright
\vspace{-0.5em}
\end{table}

Table~\ref{tab:tpl-comparison} compares {\tool} with three representative baselines in terms of both detection effectiveness and runtime efficiency. Overall, {\tool} achieves the best F1 scores across all app variants, with practical per-pair detection time.

\textbf{Effectiveness.}
On the D8 variant, all tools achieve relatively high F1 scores, while {\tool} obtains the best performance with 94.2\% library-level and 87.4\% version-level F1 scores. This indicates that when code structure is largely preserved, both existing approaches and {\tool} can capture sufficient signals for TPL detection.

As transformations become stronger, the baselines suffer substantial performance degradation, especially under R8 optimization and shrinking. In particular, LibPecker and LibScan collapse on the Opt+Srk and Opt+Obf+Srk variants due to the lack of explicit modeling of R8 optimizations. LibHunter performs better by considering method inlining but remains limited by its hand-crafted opcode-centric matching and whole-library similarity aggregation under partial reuse.

The additional obfuscation in Opt+Obf+Srk removes lexical cues such as method, field, and class identifiers, which slightly affects {\tool}. However, since our contrastive learning is trained to align methods based on functional semantics rather than identifier names, {\tool} remains robust under obfuscation. As a result, {\tool} consistently achieves the best F1 scores across all transformed variants. Under the most challenging Opt+Obf+Srk setting, {\tool} achieves 81.3\% library-level and 47.6\% version-level F1 scores, outperforming the best baseline by 32.0 and 12.5 percentage points, respectively.

These results demonstrate that context-aware contrastive method representations and functional partition-based scoring improve robustness against both code optimization and partial TPL reuse. Nevertheless, version-level detection remains more challenging than library-level detection, particularly under strong code transformations, indicating that fine-grained version discrimination is still an open problem.

\textbf{Efficiency.} For runtime comparison, we report the end-to-end online inference time amortized per app--TPL pair. We exclude only offline preprocessing of the TPL corpus, such as TPL-side feature extraction, embedding generation, and functional partitioning, which can be cached before deployment. In contrast, all target-app processing costs, including decompilation, feature extraction, candidate filtering, app-side embedding generation, similarity computation, and library-level scoring, are included. Detection is performed in a $1$-to-$N$ pipeline, where each target app is processed once and matched against $N$ candidate TPLs in parallel. We report the total online time amortized per app--TPL pair and fix the number of CPU processes to 12 for all tools.

As shown in Table~\ref{tab:tpl-comparison}, LibScan is the fastest tool due to its multi-stage early stopping strategy, while LibPecker incurs the highest overhead because it lacks a pre-filtering mechanism. Although {\tool} uses deep learning models for semantic matching, it keeps runtime practical by caching TPL-side embeddings and partitions, computing app-side embeddings only once, and reusing them across parallel matching tasks. Overall, {\tool} achieves runtime comparable to LibHunter, requiring 2.193s per pair on D8 and 0.644s per pair on Opt+Obf+Srk, while substantially improving detection effectiveness.

\begin{rqanswerbox}
\noindent\textbf{Answer to RQ1.}
{\tool} achieves the best detection effectiveness among all evaluated baselines across all app variants, especially under aggressive R8 transformations. Meanwhile, its end-to-end online inference time remains practical and comparable to LibHunter.
\end{rqanswerbox}

\subsection{Ablation Study (RQ2)}

In this RQ, we evaluate how different components and pretrained backbones affect {\tool}'s performance. Since all tools perform well on D8, we focus on the three R8-transformed variants.

\newcolumntype{C}[1]{>{\centering\arraybackslash}m{#1}}

\begin{table}[htbp]
    \renewcommand{\arraystretch}{1.08}
    \setlength{\extrarowheight}{1pt}
    \caption{F1-score for ablation study of {\tool}. Each result is shown as library-level / version-level in percentages.} %
    \label{tab:ablation-study}
    \vspace{-0.5em}
    \centering
    \begin{tabular}{C{0.14\columnwidth}|C{0.01\columnwidth}C{0.01\columnwidth}C{0.01\columnwidth}C{0.01\columnwidth}|c|c|c}
        \specialrule{1pt}{0pt}{0pt}
        
        \multicolumn{1}{c|}{\multirow{2}{*}{Backbone}} & 
        \multicolumn{4}{c|}{Configuration} &
        \multicolumn{1}{c|}{\multirow{2}{*}{Srk}} & 
        \multicolumn{1}{c|}{\multirow{2}{*}{Opt+Srk}} &
        \multicolumn{1}{@{\hspace{1.5pt}}c@{\hspace{1.5pt}}}{\multirow{2}{*}{Opt+Obf+Srk}} \\
        \cline{2-5} 
        \multicolumn{1}{c|}{} & ft & cr & cc & fp &  &  & \\ 
        \hline
        \multirow{5}{*}{UniXcoder}
& \checkmark & \checkmark & \checkmark & \checkmark & 86.0 / \textbf{65.9} & 82.4 / \textbf{50.2} & \textbf{81.3} / \textbf{47.6}  \\
        &  &  &  & \checkmark & 71.0 / 43.9 & 52.1 / 21.8 & 42.6 / 7.9  \\
        & \checkmark &  &  & \checkmark & 78.1 / 55.7 & 74.7 / 42.5 & 72.4 / 41.2 \\
        & \checkmark & \checkmark &  & \checkmark & 81.1 / 60.4 & 76.9 / 43.9 & 75.8 / 44.8 \\
        & \checkmark & \checkmark & \checkmark & & 76.2 / 62.0 & 67.3 / 43.4 & 66.7 / 43.5 \\
        \hline
        \multirow{2}{*}{CodeBERT}
        & \checkmark & \checkmark & \checkmark & \checkmark & 83.1 / 62.2 & 81.4 / 46.6 & 79.8 / 44.4  \\
        &  &  &  & \checkmark & 58.0 / 30.1 & 33.9 / 8.6 & 25.1 / 7.0  \\
        \hline
        \multirow{2}{*}{Jina-v2}
        & \checkmark & \checkmark & \checkmark & \checkmark & \textbf{86.4} / 64.9 & \textbf{82.5} / 49.3 & 80.6 / 47.0  \\
        &  &  &  & \checkmark & 70.0 / 47.0 & 70.2 / 31.5 & 55.9 / 15.3 \\

        \specialrule{1pt}{0pt}{0pt}
    \end{tabular}
\vspace{-0.3em}
\end{table}

\textbf{Contribution of Different Components.} 
To study the contribution of different design components, we fix the base model to the default UniXcoder used in our approach, and selectively enable or disable the following four options:
\begin{itemize}
\item \textbf{Fine-tuning (ft)}: When disabled, the encoder uses the off-the-shelf pre-trained base model, allowing us to quantify the impact of task-specific representation adaptation.
\item \textbf{Call Relationship (cr)}: When disabled, the \textbf{\cgencoder} encodes only the target method itself, allowing us to assess the contribution of call relationship information.
\item \textbf{Class Context (cc)}: When disabled, the \textit{Contrastive Method Embedding Model} reduces from a dual-encoder architecture to a single-encoder one based solely on the \textbf{\cgencoder}. This removes the \textbf{\ctxencoder} and the associated cross-attention module (\textbf{\crossattn}), allowing us to isolate and assess the contribution of class-level contextual information.
\item \textbf{Functional Partitioning (fp)}: When disabled, the library-level similarity score is computed directly using Equation~\ref{eq:vsim} adopted by existing tools~\cite{libhunter,libscan}, allowing us to assess the impact of the proposed functional partition-based scoring strategy.
\end{itemize}

Each model architectural variant is fine-tuned separately using the same contrastive learning objective, ensuring that performance differences are attributable to architectural choices rather than parameter reuse.

Table~\ref{tab:ablation-study} reports the ablation results. Contrastive fine-tuning consistently improves performance over the off-the-shelf UniXcoder, especially under the most challenging Opt+Obf+Srk setting, showing that task-specific training helps the encoder learn transformation-resilient semantic representations. Adding call relationship and class context further improves both library-level and version-level F1 scores, confirming that contextual cues complement isolated method representations. Finally, disabling functional partitioning leads to clear performance drops across all R8 variants, demonstrating its importance for handling partial TPL reuse.

\textbf{Effect of Pretrained Model Choice.}
To assess whether {\tool} depends on a specific pretrained backbone, we replace the default UniXcoder~\cite{unixcoder} with two widely used alternatives, CodeBERT~\cite{feng2020codebert} and Jina-embeddings-v2-base-code (Jina-v2)~\cite{jina-embedding-v2-base-code}. These code representation models have been broadly adopted in software engineering tasks, including code search and automated program repair~\cite{xia2022less,huang2023empirical-repair,xia2023automated,li2024programming,wiedemeier2025walking,shi2025large}.

With functional partitioning enabled, we evaluate each backbone in two settings: (1) using only the pretrained encoder without contextual information, and (2) integrating it into the full {\tool} pipeline with the same context-aware contrastive fine-tuning objective.

As shown in Table~\ref{tab:ablation-study}, pretrained encoders alone are insufficient under R8 transformations, especially for version-level detection under Opt+Obf+Srk. After incorporating contextual semantics and contrastive fine-tuning, all three backbones achieve substantial improvements, indicating that the proposed training and context modeling strategy is effective across different pretrained models. UniXcoder and Jina-v2 obtain comparable results, while CodeBERT is slightly weaker.

\begin{rqanswerbox}
\noindent\textbf{Answer to RQ2.}
Both components improve {\tool}: context-aware contrastive learning strengthens method matching, and functional partitioning handles partial TPL presence. Consistent gains across pretrained backbones further indicate that the improvements stem from the proposed framework rather than any specific base model.

\end{rqanswerbox}

\subsection{Sensitivity (RQ3)}

In this RQ, we investigate the sensitivity of the final F1 score to the minimum number of methods required for each functional partition. We evaluate the performance of {\tool} on the Srk, Opt+Srk, and Opt+Obf+Srk variants in the test set of {\ourdataset} by varying $\thresholdpartition$ from 100 to 1000 with a step size of 100.

\begin{figure}[htbp]
    \centering

    \begin{subfigure}[h]{1\linewidth}
     \setlength{\abovecaptionskip}{0pt} %
    \setlength{\belowcaptionskip}{0pt} %
        \centering
        \includegraphics[width=\linewidth]{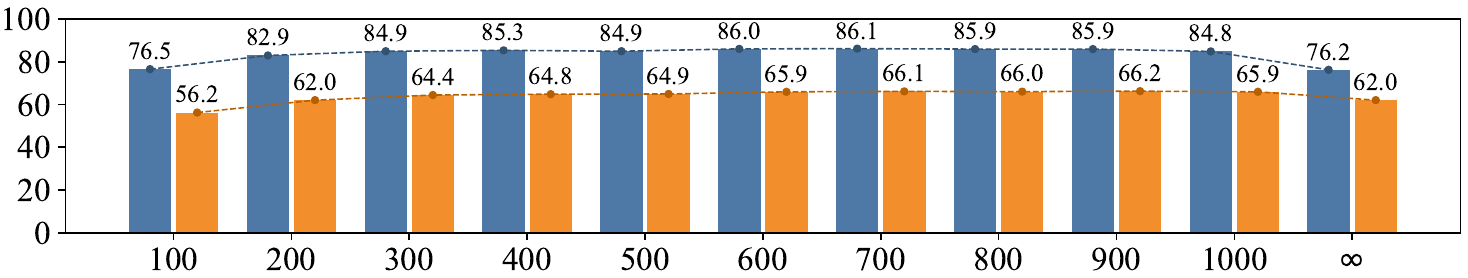}
        
        \caption{Srk}
        \label{fig:srk-sens}
    \end{subfigure}

    \begin{subfigure}[h]{1\linewidth}
     \setlength{\abovecaptionskip}{0pt} %
    \setlength{\belowcaptionskip}{0pt} %
        \centering
        \includegraphics[width=\linewidth]{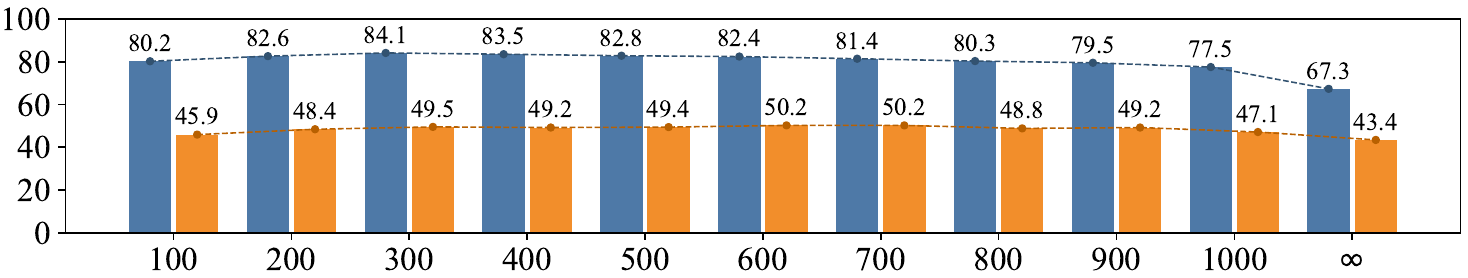}
        \caption{Opt+Srk}
        \label{fig:opt-srk-sens}
    \end{subfigure}

    \begin{subfigure}[h]{1\linewidth}
     \setlength{\abovecaptionskip}{0pt} %
    \setlength{\belowcaptionskip}{0pt} %
        \centering
        \includegraphics[width=\linewidth]{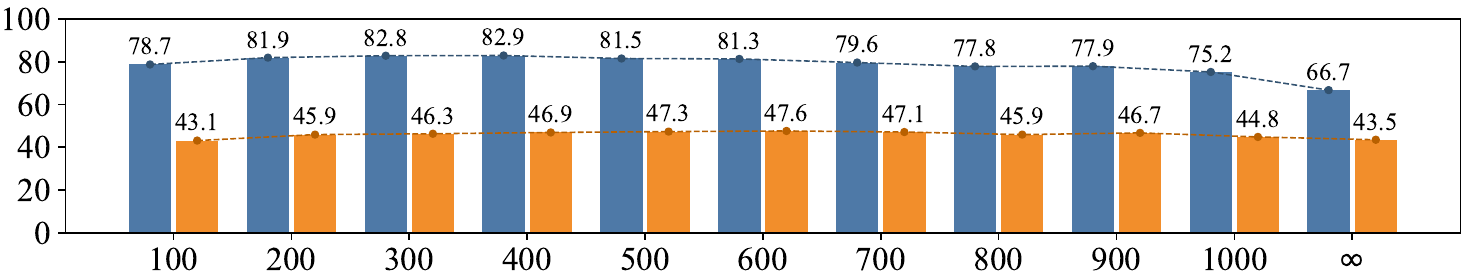}
        \caption{Opt+Obf+Srk}
        \label{fig:opt-obf-srk-sens}
    \end{subfigure}

    \begin{subfigure}[h]{1\linewidth}
        \centering
        \includegraphics[width=\linewidth]{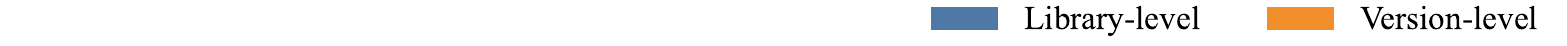}
    \end{subfigure}
    \caption{F1-score Sensitivity to the functional partition threshold}
    \label{fig:sensitivity}
    \vspace{-0.5em}
\end{figure}

Figure~\ref{fig:srk-sens}, Figure~\ref{fig:opt-srk-sens}, and Figure~\ref{fig:opt-obf-srk-sens} show the sensitivity of {\tool} to different values of $\thresholdpartition$ on the three R8-transformed variants, where $\infty$ denotes disabling functional partitioning. Overall, {\tool} remains stable when $\thresholdpartition$ ranges from 200 to 900, with only minor fluctuations at both the library and version levels. On the most challenging Opt+Obf+Srk variant, the best library-level and version-level F1 scores are achieved at $\thresholdpartition=400$ and $\thresholdpartition=600$, respectively.

Very small partitions, such as $\thresholdpartition=100$, lead to noticeable performance drops because method-level false positives can be amplified when the denominator of a partition is too small. Conversely, overly large partitions, such as $\thresholdpartition=1000$, reduce the benefit of partitioning and gradually converge to the non-partitioned setting.

\begin{rqanswerbox}
\noindent\textbf{Answer to RQ3.}
{\tool} remains stable across a broad range of $\thresholdpartition$ values. Moderate partition sizes work best, avoiding false positives from overly small partitions and the loss of partitioning benefits from overly large ones.
\end{rqanswerbox}

\subsection{Generalizability (RQ4)}

In this RQ, we evaluate whether {\tool} generalizes beyond the training setting by testing it on unseen TPLs and previously unseen real-world in-the-wild apps.

\textbf{Unseen TPLs.}
To assess whether {\tool} generalizes to TPL artifacts not observed during contrastive fine-tuning or checkpoint selection, we construct a TPL-disjoint subset of {\ourdataset}, where test TPLs do not overlap with those in the training or validation sets. Since real-world TPL usage follows a long-tailed distribution, directly constructing such a split is difficult: rare TPLs provide too few app--TPL instances for reliable evaluation, whereas highly popular TPLs appear in many apps and create extensive overlap conflicts across subsets, leaving too few apps after enforcing TPL disjointness. Inspired by prior practices for handling long-tailed and sparse observations~\cite{liu2019large,mikolov2013distributed,hellendoorn2017deep}, we retain TPLs appearing 5--50 times among the 200 apps in {\ourdataset}. The lower bound filters out extremely sparse TPLs, while the upper bound excludes overly shared TPLs to preserve enough apps for the split.

The resulting split contains 17 training apps with 25 app--TPL pairs and 18 validation apps with 25 app--TPL pairs. The training and validation sets share the same 9 seen TPL artifacts with 60 versions, which are used for model fine-tuning and checkpoint selection. The test set contains 61 apps with 83 app--TPL pairs, covering 10 held-out TPL artifacts with 77 versions.
We keep the RQ1 hyperparameters unchanged without subset-specific tuning.

\begin{table}[htbp]
\renewcommand{\arraystretch}{1.08}
\setlength{\extrarowheight}{1pt}
\caption{Library-level / version-level F1 scores of {\tool} and LibHunter on the TPL-disjoint test set.}
\label{tab:unseen-tpl-comparison}
\vspace{-0.5em}
\centering
\resizebox{0.8\columnwidth}{!}{
\begin{tabular}{c|c|c|@{\hspace{2pt}}c@{\hspace{2pt}}}
\specialrule{1pt}{0pt}{0pt}
Tool & Srk & Opt+Srk & Opt+Obf+Srk \\
\hline
LibHunter & 67.8 / 58.7 & 54.9 / 42.0 & 55.5 / 42.7 \\
\hline
{\tool} & \textbf{81.0} / \textbf{69.3} & \textbf{78.9} / \textbf{54.4} & \textbf{78.3} / \textbf{53.8} \\
\specialrule{1pt}{0pt}{0pt}
\end{tabular}}
\vspace{-0.3em}
\end{table}

As shown in Table~\ref{tab:unseen-tpl-comparison}, {\tool} consistently outperforms LibHunter across all three R8 settings on the test set. Under the most challenging Opt+Obf+Srk setting, {\tool} achieves 78.3\%/53.8\% library-level/version-level F1, outperforming LibHunter by 22.8/11.1 percentage points. These results provide evidence that {\tool}'s improvements are not limited to TPL artifacts observed during contrastive fine-tuning or checkpoint selection.

\textbf{In-the-wild Apps.}
To assess whether {\tool} can surface security-relevant dependency signals in real-world deployment, we further apply {\tool} to 1,081 previously unseen Google Play apps, collected from the top 30 apps in each of 37 categories after removing duplicate package names. We use the 46 CVE-associated TPLs from {\ourdataset} as the candidate set and also run LibHunter as a reference baseline on the same apps.

{\tool} reports 9,102 app--TPL pairs, corresponding to 8.42 candidate TPLs per app on average, while LibHunter reports 4,460 pairs.
Among {\tool}'s reports, 3,478 are associated with TPL versions that have historical CVE records, accounting for 38.2\% of all reported pairs, while LibHunter reports 1,392 CVE-associated version reports, accounting for 31.2\% of all its reported pairs.
These CVE-associated reports from {\tool} serve as potential vulnerable-version warnings for downstream dependency risk analysis. Such warnings frequently involve transitive TPLs such as \texttt{protobuf-\allowbreak java}, \texttt{okio}, \texttt{gson}, and \texttt{guava}, suggesting that {\tool} can help surface hidden dependency risks in real-world apps.

\begin{rqanswerbox}
\noindent\textbf{Answer to RQ4.}
{\tool} generalizes to TPLs unseen during fine-tuning or checkpoint selection, consistently outperforming LibHunter on the TPL-disjoint test set. On in-the-wild Google Play apps, it also surfaces potential security-relevant dependency signals at scale, demonstrating practical applicability beyond curated benchmarks.
\end{rqanswerbox}

\section{Discussion}

\textbf{Limitations.} Although our approach achieves notable improvements in version-level detection, the overall accuracy at this level remains suboptimal.
This limitation mainly stems from the subtle differences between adjacent TPL versions, which are often difficult to capture using global similarity-based matching, especially under compiler optimizations that may obscure or eliminate version-specific changes.
We believe that improving version-level detection requires a more fine-grained analysis of version-differentiating code segments.
A promising direction for future work is to decouple TPL detection into two stages: (1) reliably identifying the presence of a TPL, and (2) pinpointing the specific version by focusing on code regions that differentiate adjacent versions and remain stable across code transformations.

\textbf{Threats to Internal Validity.}
Our approach relies on Androguard for call graph construction and contextual information extraction. While widely adopted, Androguard may produce incomplete call graphs due to its limited handling of advanced language features such as polymorphism, reflection, and multi-threading. In addition, our analysis operates on Java code decompiled by JADX, whose decompilation quality may be imperfect in the presence of aggressive compiler transformations. Nevertheless, these tools are widely used in prior work, and their limitations apply uniformly across all evaluated approaches, which helps mitigate potential bias in comparative evaluation.

\textbf{Threats to External Validity.}
Although we construct a dataset of 200 real-world apps compiled under various R8 settings, these apps are primarily sourced from the open-source F-Droid ecosystem. As a result, the app distribution may differ from that of commercial app markets, where proprietary development practices, closed-source libraries, and additional protection mechanisms are more prevalent.
Establishing accurate ground truth for third-party library usage in such settings remains a challenging open problem and warrants further investigation.

\section{Related Work}

\textbf{Android TPL Detection.} Many approaches have been proposed for Android TPL
detection~\cite{ma2016libradar,zhang2019libid,pan2025pay}.
LibScout~\cite{libscout} uses class hierarchy features and Merkle tree matching to identify TPLs and their versions.
LibD~\cite{li2017libd} detects Android TPLs at scale by analyzing package dependency relations and clustering library code across apps.
LibPecker~\cite{libpecker} constructs class-level signatures with weighted similarity scoring to tolerate code customization. Orlis~\cite{orlis} employs inter-procedural structural features and similarity digests for efficient matching.
ATVHunter~\cite{zhan2021atvhunter} integrates control-flow and opcode-level analysis to detect vulnerable TPL versions accurately.
LibScan~\cite{libscan} first matches class-level signatures and then compares method-level opcodes, further incorporating call-chain opcode similarity to improve robustness against obfuscation.
LibHunter~\cite{libhunter} explicitly considers code optimization. It designs a more generalized class signature matching mechanism to counter call site optimization, and simulates R8’s method inlining strategy at the method level to enhance method matching, thereby resisting method inlining.
However, existing approaches still provide limited support for high-level code semantics and partial TPL reuse, reducing their effectiveness in complex TPL detection scenarios.

\textbf{Learning-based Binary Code Similarity Detection.} To address code transformations introduced by compilers, extensive learning-based approaches have been proposed, primarily in the domain of \textit{Binary Code Similarity Detection (BCSD)}~\cite{li2019gmn,xu2017gemini,ding2019asm2vec,massarelli2019safe,wang2022jtrans}. 
For instance, HermesSim introduces a Semantics-Oriented Graph to characterize data flow, control flow, and side-effect features, thereby improving the robustness of BCSD~\cite{he2024hermersim}. Furthermore, in the domain of \textit{Binary Software Composition Analysis}, LibDB uses neural function embeddings and call graph matching for native binary TPL reuse detection~\cite{tang2022libdb}. BinaryAI proposes an embedding-based function retrieval strategy and a locality-driven matching objective to detect the reused components~\cite{jiang2024binaryai}. 
Nevertheless, these approaches are primarily designed for C/C++ binary analysis tasks and lack adaptation to Android-specific contexts (e.g., method call relationship and class-level context), rendering them unsuitable for direct application to the Android TPL detection task.

\section{Conclusion}

We propose {\tool}, a learning-based Android third-party library detection tool that realizes context-aware functional modeling for robust TPL detection under code transformations and partial reuse. {\tool} combines context-aware contrastive learning with TPL functional partitioning and outperforms the baselines at both the library and version levels across the evaluated benchmarks.

The implementation of {\tool} is available at \url{https://anonymous.4open.science/r/libfan}. The dataset {\ourdataset} and the trained models are available at \url{https://figshare.com/s/e7848208db8ceb802883}.

\bibliographystyle{IEEEtran}
\bibliography{LibFan/ref}

\end{document}